# Experimental Determination of the Spectral Function of Graphene

Aaron Bostwick,[1] Taisuke Ohta,[1,2] Thomas Seyller,[3] K. Horn[2] and Eli Rotenberg[1]

[1]*Advanced Light Source, E. O. Lawrence Berkeley National Laboratory, Berkeley, CA 94720 USA*

[2] *Department of Molecular Physics, Fritz Haber Institute, Germany*

[3] *Institut für Physik der Kondensierten Materie, Universität Erlangen-Nürnberg, Germany*

Abstract

A number of interesting properties of graphene and graphite are postulated to derive from the peculiar bandstructure of graphene. This bandstructure consists of conical electron and hole pockets that meet at a single point in momentum (k) space—the Dirac crossing, at energy $E_D = \hbar\omega_D$. Direct investigations of the accuracy of this bandstructure, the validity of the quasiparticle picture, and the influence of many-body interactions on the electronic structure have not been addressed for pure graphene by experiment to date. Using angle resolved photoelectron spectroscopy (ARPES), we find that the expected conical bands are distorted by strong electron-electron, electron-phonon, and electron-plasmon coupling effects. The band velocity at $E_F$ and the Dirac crossing energy $E_D$ are both renormalized by these many-body interactions, in analogy with mass renormalization by electron-boson coupling in ordinary metals. These results are of importance not only for graphene but also graphite and carbon nanotubes which have similar bandstructures.

With the recent discovery of superconductivity in carbon nanotubes (CNTs)[1],[2], alkaline-metal-doped $C_{60}$ crystals[3], and graphite intercalation compounds[4],[5],[6] (GICs) with relatively high transition temperatures, there is a strong interest in the influence of many-body interactions on the electron dynamics of graphite and related materials. Graphene is a sheet of carbon atoms distributed in a hexagonal lattice and is the building block for all of these materials; therefore it is a model system for this entire family.

Graphene and graphite are also interesting because their simple bandstructures reflect a novel property: their carriers can be described as relativistic Fermions through a formal equivalence of the wave equation with the relativistic Dirac equation.[7],[8],[9] That they travel as massless particles with a fixed "speed of light" c*~c/300 follows from the nearly linear dispersion of the bands through $E_D$. Departure of the electron dynamics from vFermi liquid behavior in graphite has also been attributed to the special shape of the graphene bandstructure[10],[11]

These interesting features are a consequence of graphene's half-filled valency and the simple hexagonal arrangement of its C atoms. The bandstructure $E(\mathbf{k})$ may be described by a simple one-orbital tight-binding

model as[12]

$$E(\mathbf{k}) = \pm t\sqrt{1 + 4\cos(\sqrt{3}ak_y/2)\cos(ak_x/2) + 4\cos^2(ak_x/2)} \qquad (1)$$

where $\mathbf{k}$ is the in-plane momentum, $a$ is the lattice constant, and $t$ is the near-neighbor hopping energy.

In Figure 1 we compare energy bands and constant energy surfaces computed using Equation 1 to ARPES measurements applied to a single layer of graphene grown on the (0001) surface of SiC (6H polytype). The primary bands, cones centered at the K points, are surrounded by six weak replica bands; these result from the interference of the graphite and substrate lattice constants (2.4 vs. 3.07Å) and correspond to similar satellite spots in low energy electron diffraction.[13] The primary bands are in good overall agreement with the simple model despite its having only two adjustable parameters: the hopping energy $t = 2.82$ eV and a 0.435 eV shift of the Fermi energy $E_F$ above the Dirac crossing energy $E_D$. This shift is attributed to doping of the graphene layer by depletion of the substrate's *n*-type carriers near the SiC surface.

Despite this good agreement, profound deviations are observed near $E_F$ and $E_D$. We show in Figure 2a a magnified view of the bands measured along a line (the vertical double arrow in Figure 1b) through the K point. The predicted, or "bare" bands in this direction are nearly perfectly linear and mirror-symmetric with respect to the K point according to Equation 1. The actual bands deviate from this prediction in two significant ways: first, at a binding energy $\hbar\omega_{ph}=200$ meV below $E_F$, we observe a sharpening of the bands accompanied by a slight kink in the bands' dispersions. We attribute this feature to renormalization of the electron bands near $E_F$ by coupling to phonons, as discussed later.

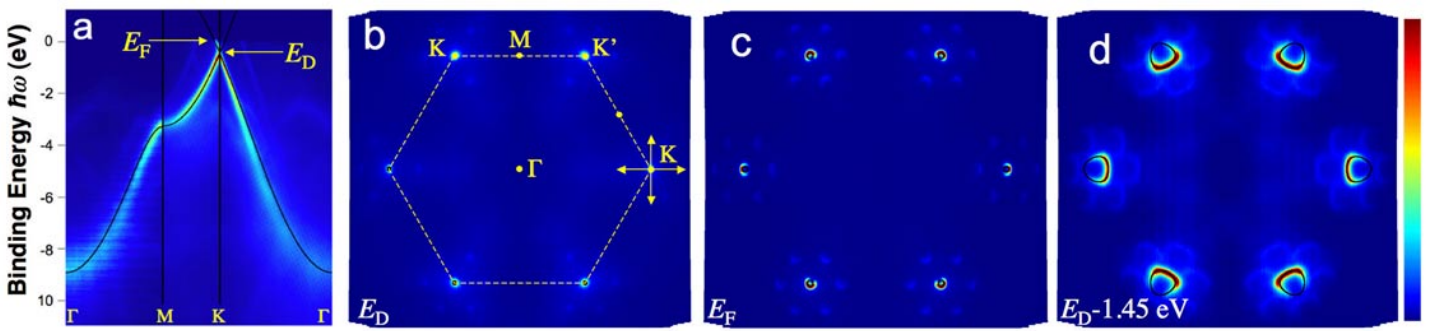

**Figure 1 | The bandstructure of graphene.** **a** The experimental energy distribution of states as a function of momentum along principal directions, together with a single-orbital model (solid lines) given by Eq. (1). **b** Constant energy map of the states at binding energy corresponding to the Dirac energy ($E_D$) together with the Brillouin Zone boundary (dashed line). The orthogonal double arrows indicate the 2 directions over which the data in Fig. 2 were acquired. **c-d** Constant energy maps at the Fermi energy ($E_F = E_D + 0.45$) and $E_D - 1.5$ eV, respectively. The faint replica bands correspond to the 6√3 × 6√3 satellite peaks in low energy electron diffraction.[13]

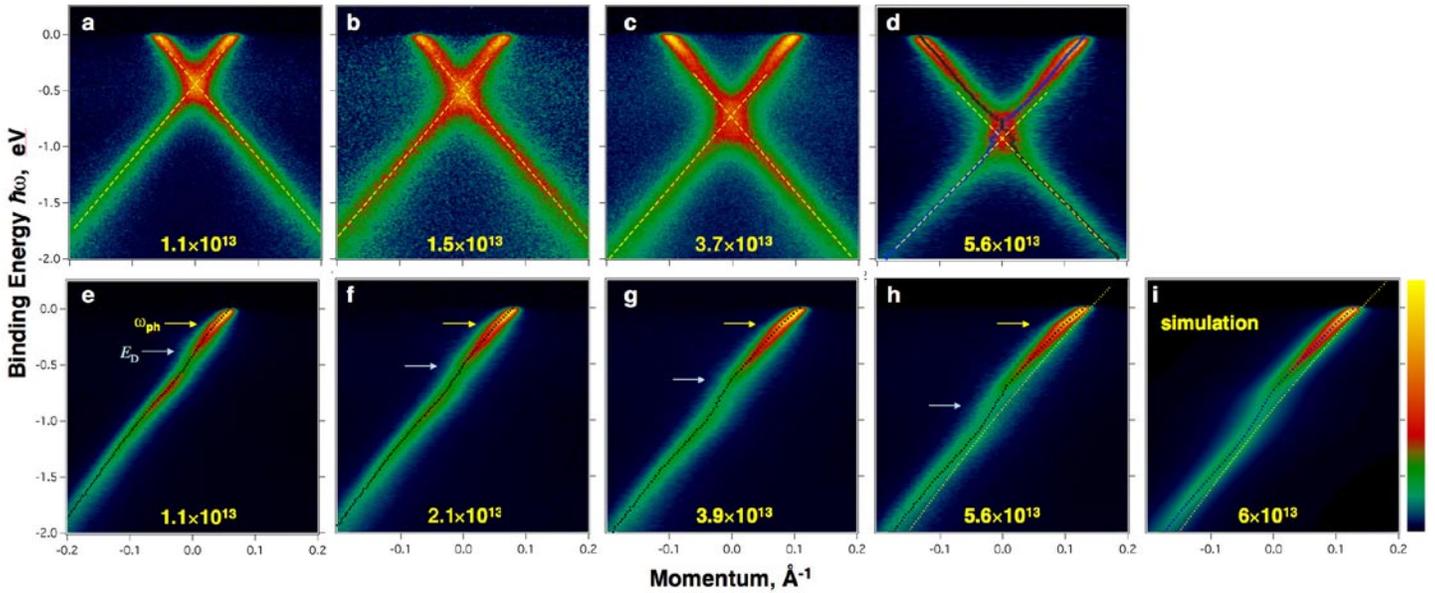

**Figure 2 | The bandstructure of graphene near the Fermi Level.** a-d Experimental energy bands along a line through the K point parallel to GM direction (along the vertical double-arrow in Figure 1b) as a function of progressively increased doping by potassium adsorption. The dashed lines are an extrapolation of the lower bands (below the Dirac crossing energy $E_D$), which are observed not to pass through the upper bands (above $E_D$), suggesting the kinked shape of the bands around $E_D$. The electron density (per cm$^2$) is indicated on each panel. e-h Bandmaps for similar dopings acquired in an orthogonal direction through the K point (horizontal double arrow in Figure 1b), for which one of the bands is suppressed. The non-linear, or "kinked" dispersion of the bands together with linewidth variations (corresponding to Re and Im parts of the self energy S) are clearly visible. The kinks, marked by arrows, occur at a fixed 200 meV energy and near $E_D$, the latter varying with doping. i The simulated spectral function, calculated using only the bare band (yellow dotted line) and Im$\Sigma$ derived from the data in panel h.

Second, linear extrapolations of the lower bands (dashed lines in Figure 2a) do not pass through the upper bands, demonstrating that the bands do not pass smoothly through $E_D$ as Equation 1 predicts. This is observed more easily for data acquired along the orthogonal direction through the K point (Figure 2e), because, first a matrix element effect[14] suppresses one of the two bands and second, this cut is perpendicular to the Fermi contour. We find that the near $E_D$ the bands have an additional kink, which we propose is caused by additional many-body interactions.

The deviations from the bare band are sensitive to doping, which we varied by adsorbing potassium atoms that readily donate electrons to the film. The evolution of the band structure upon doping is followed in Figure 2b-d and along the orthogonal direction in Figure 2f-h. Similar to graphite, doping graphene by K deposition shifts the bands more or less rigidly to higher binding energy.[15] While the energy of the kink at 200 meV does not change, the second kink strengthens and follows $E_D$ with doping, demonstrating that it is associated with electrons with energy near $E_D$. The effect of this kink on the bandstructure is significant: at high doping, a curve fit of the band positions (Figure 2d) shows that $E_D$ has been shifted towards $E_F$ by 130 meV from the single-particle prediction.

In the quasiparticle scheme, the carriers are represented as single particles that scatter from and are surrounded by a cloud of other 'particles' (such as phonons); the entire entity moves somewhat like a free particle but with renormalized energy. In this scheme, ARPES measures the spectral function, expressed in terms of the complex self energy $\Sigma(\mathbf{k},\omega)$, as

$$A(\mathbf{k},\omega) = \frac{|\mathrm{Im}\,\Sigma(\mathbf{k},\omega)|}{(\omega - \omega_b(\mathbf{k}) - \mathrm{Re}\,\Sigma(\mathbf{k},\omega))^2 + (\mathrm{Im}\,\Sigma(\mathbf{k},\omega))^2} \qquad (2)$$

where $\mathbf{k}$ and $\omega$ are the quasiparticle momentum and energy, and $\omega_b$ is the bare band dispersion in the absence of many-body effects. $\Sigma(\mathbf{k},\omega)$ encodes both the scattering rate and the renormalization of the band dispersion in its imaginary and real parts, respectively. In the $\mathbf{k}$-independent approximation[16,17] ($\Sigma(\mathbf{k},\omega) \approx \Sigma(\omega)$), $\mathrm{Im}\Sigma(\omega)$ is proportional to the Lorentzian linewidth of the momentum distribution curve (MDC) $A(\mathbf{k},\omega)$ taken at constant $\omega$. $\mathrm{Re}\Sigma(\omega)$ is readily computed from $\mathrm{Im}\Sigma(\omega)$ through a Hilbert transform (to satisfy causality), and the full spectral function $A(\mathbf{k},\omega)$ can be reconstructed using the computed $\mathrm{Re}\Sigma(\omega)$ and compared to experiment. Such a reconstruction for one doping is shown in Figure 2i; it is in excellent agreement with the data (Figure 2h) from which $\mathrm{Im}\Sigma$ was obtained. This shows that the kinks in the bands originate not from details of the single-particle bandstructure, but rather from many-body interactions, providing strong support for the quasiparticle picture in graphene.

The observed kink structure is therefore derived from a complicated $\omega$-dependence of the observed scattering rate (proportional to the MDC linewidths shown in Figure 3) as a function of doping. To model this behavior, we consider three processes: decay of the carriers by phonon emission, by electron-hole (*e-h*) pair generation, and by emission of collective charge excitations (plasmons) via electron-plasmon (*e-pl*) coupling. (Impurity scattering, a fourth contributing process, can be neglected as its contribution to the MDC linewidth is smaller than the experimental momentum resolution (~0.01Å$^{-1}$) and in any case merely leads to a uniform background scattering rate). By summing up all the momentum- and energy-conserving decay events as a function of hole energy $\omega$, we can show that the three principle decay processes contribute differently to the lifetime in regions I-IV as identified in Figure 3.

Such a calculation for the total scattering rate, together with the individual contributions from *e-ph*, *e-h* and *e-pl* processes, is shown in Figure 3. This calculation is for a sample with $n=5.6\times10^{13}$ cm$^{-2}$ and compares favorably to experimental MDC width for that doping; similar agreement can be obtained for the other dopings as well. The predicted dip at $E_D$ is an artifact of the simplicity of our model, which does not consider interactions between the plasmons and the Fermi liquid excitations.

Now we discuss the different decay processes in turn. We attribute the kink near $E_F$ to electron-phonon (*e-ph*) coupling as described previously for metals[18,19,20], and (possibly) high-$T_c$ superconductors[21,22]. *E-ph* coupling is expected at this energy scale considering the known phonon spectrum.[23] In this process, photoholes decay by phonon emission (see Figure 4a). Using the graphite phonon density of states[24], we calculated the *e-ph* contribution to ImΣ (Figure 3, green curve) with the standard formalism[25] and find an *e-ph* coupling constant $\lambda \approx 0.3$. Although this is a factor of 5 larger than predicted[26] for $n=5.6\times10^{13}$ cm$^{-2}$, comparison with the experimental data shows that this provides an accurate description of ImΣ in region I. The observed increase of the kink's strength with $n$ (see Figure 2e-h) is expected from the increase in the size of the Fermi surface, although the 200 meV energy scale remains constant because the K atoms should not alter the phonon bandstructure.

Consider now the decay of the photohole by excitation of an electron from below to above $E_F$, thereby creating an *e-h* pair. In Landau's model of the Fermi liquid (FL), the scattering rate from such processes increases as $\sim\omega^2$ away from $\omega=0$, reflecting the growing number of possible excitations that satisfy momentum and energy conservation. However, the linear dispersion of the graphene bands and the presence of the Dirac crossing drastically modify this picture. A hole just above $E_D$ can easily decay through many possible *e-h* creation events, for example as in Figure 4b, and we find a similar $\omega^\alpha$ ($\alpha\sim1.5$) dependence of ImΣ in regions I-II as in FL theory. But a hole decay originating at $\omega$ just below $E_D$ has few final states with sufficient momentum transfer to excite an *e-h* pair (Figure 4c). This causes a sharp reduction in the scattering rate in region III. Only for energies $\omega \lesssim 2\omega_D$, region IV, does *e-h* pair generation become favorable (e.g. Figure 4d).

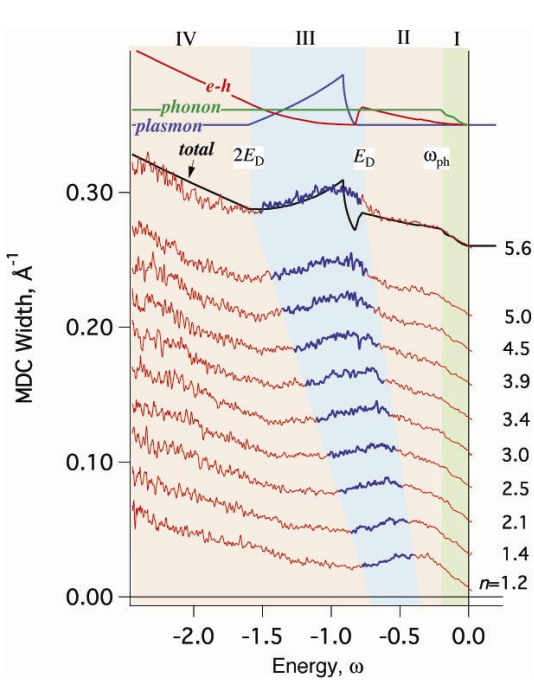

**Figure 3 | MDC Widths of Carriers in Graphene.** Measured spectral MDC width (assumed proportional to scattering rate and ImΣ) for graphene, derived by performing a line shape analysis of momentum distribution curves for each binding energy as a function of doping $n$ (in units $10^{13}$ cm$^{-2}$). Each trace is shifted upward by 0.025Å$^{-1}$. The simulated total scattering rate (black line) and the partial contributions due to decay into phonons (green), electron-hole pairs (red) and plasmons (blue) are compared to the MDC spectral width for the highest doped sample. These interactions contribute differently in regions I-IV defined as follows: (I) the phonon energy scale $\omega_{ph} < \omega < 0$, (II) the Dirac energy scale $\omega_D < \omega < 0$, (III) $2\omega_D < \omega < \omega_D$, (IV) $\omega \lesssim 2\omega_D$.

The *e-h* and *e-ph* processes can explain the observed MDC widths in regions I, II, and IV. In region III, however, decay by *e-h* pair creation is practically not allowed yet the observed scattering rate has a peak rather than a dip (highlighted in blue in Figure 3). We now show that this peak may be explained by decay through plasmon emission. Plasmons are oscillations of an electron gas that play an important role in the optical properties of ordinary metals. In graphene, the charge carriers near the K point have zero effective mass and travel like photons at constant speed $c^*$, but unlike photons, they have charge and are therefore subject to collective oscillations such as plasmons. Although a full treatment of the *e-pl* interaction is difficult near the Dirac point, a simple model suffices to explain how *e-pl* coupling can enhance the scattering rate below $E_D$.

Ordinary two-dimensional plasmons have a dispersion relationship

$$\omega_{pl}(q) = \sqrt{4\pi n e^2 q / m(1+\varepsilon)} \qquad (3)$$

where $q$ is the plasmon momentum, $m$ is the carrier mass, and $\varepsilon \sim 10$ is the dielectric constant. Although plasmons in principle exist in the domain $0 < q < \infty$, in practice they propagate freely up to a critical momentum $q < q_c$ due to Landau damping (plasmon decay into electron-hole pairs)[27].

For graphene, the rest mass $m$ is zero near $E_D$, but the "relativistic mass" $m_r = E/c^{*2}$ is on the order[8] of $0.1 m_e$ and can be used to set the plasmon energy scale $\omega_{pl}$. Since the plasmon spectrum $\omega_{pl}(q)$ rises steeply over a short, finite range of $q$, decay of the photohole into plasmons becomes kinematically possible only for hole decays originating just below $E_D$ (Figure 4e).

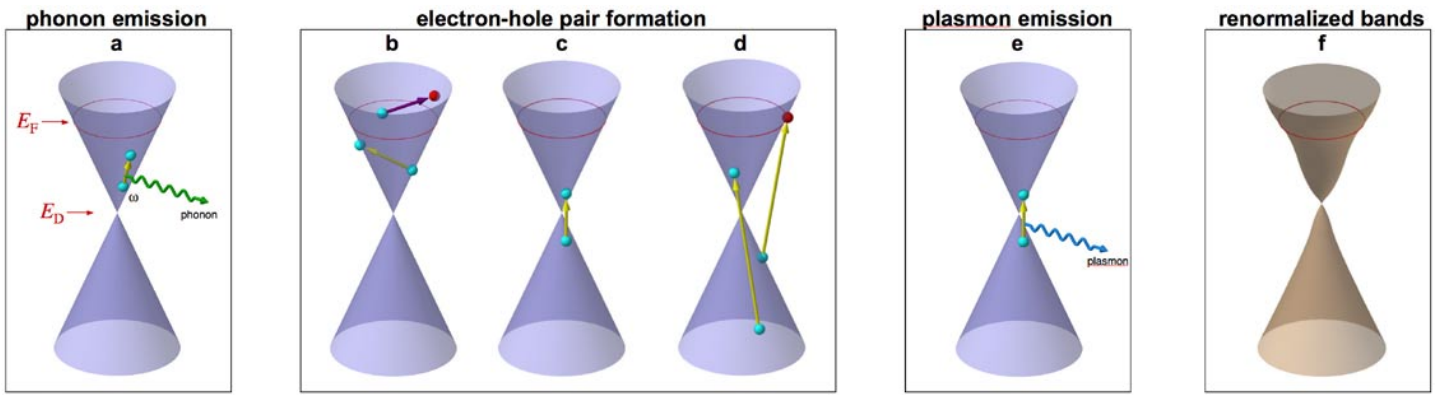

**Figure 4 | Decay and Scattering Processes in Graphene. a** The energy/momentum diagrams for decay processes scattering of a photohole, initially created at energy ω, decaying by a emission of a phonon. **b-d** Spontaneous generation of an electron-hole pair near the Fermi level $E_F$ for photohole energy satisfying **b** ($\omega > \omega_D$), **c** ($2\omega_D < \omega < \omega_D$), which can generate no possible *e-h* pair as drawn, and **d** ($\omega < 2\omega_D$). **e** emission of a plasmon ($2\omega_D < \omega < \omega_D$). **f** The net effect of these processes is to distort the bare bands to the renormalized bands (shown in tan).

Given the plasmon dispersion relation we can easily sum up the possible plasmon decays as a function of ω (Figure 3, blue curve), which is proportional to the scattering rate. We find a peak located just below $E_D$, whose width and intensity scales with $E_D$. A peak following these trends is clearly observed in the experimental data (highlighted in blue in Figure 3).

Previously, *e-pl* coupling was shown to affect the unoccupied bands of a 3-dimensional metal at the large plasmon energy scale (~20 eV)[28] But *e-pl* coupling at small energy scales is normally forbidden for two- and three-dimensional electron gases (except for the special case of layered electron gases)[10] so this is a unique instance where *e-pl* coupling is kinematically allowed for a pure 2D system. It is also remarkable because of the relativistic nature of the carriers and the strong role it plays in shifting the Dirac energy.

It is worth emphasizing that the model for the scattering rate has only four adjustable scaling factors: the *e-ph* coupling constant λ, the absolute probabilities for *e-h* pair creation and plasmon emission, and the screening constant ε which scales the Coulomb interaction. The other inputs are the experimentally determined band dispersion, the graphite phonon density of states, and the relativistic mass $m_r$ which is taken from the literature.[8]

These results show that the special condition of massless Dirac Fermions found in graphene does not preclude the validity of the quasiparticle picture – in fact the quasiparticle picture is valid over a spectacularly wide energy range – but it does induce novel *e-h* and *e-pl* decay processes. These result in strong modifications of the band dispersion, as schematically illustrated in Figure 4f. This distortion occurs not only near the Fermi level as in conventional metals, but also centered around the Dirac crossing energy $E_D$. The effects we describe are not unique to high doping levels, but extrapolate all the way down to zero doping. Near this regime (already approached for the lowest dopings in Figure 3b), the energy scales for *e-h*, *e-pl*, and *e-ph* decay processes overlap, and a unified treatment of all these interactions is necessary to reproduce the many-body effects. These conclusions apply as well to graphite, nanotubes, and other carbon materials with similar electronic structure.

Acknowledgements

This work and the ALS were supported by the Department of Energy, Office of Basic Sciences. K. H. and T. O. were supported by the Max Planck Society. A. B. and T. O. contributed equally to the measurements and authorship.

Methods.

The ARPES measurements were conducted at the Electronic Structure Factory endstation at beamline 7.01 at the Advanced Light Source, equipped with a hemispherical Scienta R4000 electron analyzer. The single layer of graphene was prepared by etchin a 6H-SiC(0001) substrate (*n*-type with a nitrogen concentration of $5\times10^{18}$ cm$^{-3}$) in a hydrogen plasma followed by annealing at 1150C for 4 minutes by direct current heating in pressure better than $1\times10^{-10}$ Torr[13]. Measurements were conducted at a pressure better than $2.5\times10^{-11}$ Torr with the sample cooled to ~20K using a photon energy $h\nu$ of 95 eV and with an overall energy resolution of ~25 meV. Potassium deposition was by a commercial (SAES) getter source. The potassium coverage can be estimated from the carrier density assuming a charge transfer of 0.7e$^-$ per alkali atom[29] to be about 0.007 monolayers when $n=1\times10^{13}$ cm$^{-2}$.

References

[1] Tang, Z. K. *et al.* Superconductivity in 4 Angstrom Single-Walled Carbon Nanotubes. *Science* **292**, 2462-2465 (2001).

[2] Kociak, M. *et al.* Superconductivity in Ropes of Single-Walled Carbon Nanotubes. *Phys. Rev. Lett*. **86**, 2416–2419 (2001).

[3] Hebard, A. F. *et al.* Superconductivity at 18 K in potassium-doped C60. *Nature* **350**, 600 – 601 (1991).

[4] Hannay, N. B. *et al.* Superconductivity in graphitic compounds. *Phys. Rev. Lett.* **14**, 225-226 (1965).

[5] Weller, T. E., Ellerby, M., Saxena, S. S., Smith, R. P. & Skipper, N. T. Superconductivity in the intercalated graphite compounds C$_6$Yb and C$_6$Ca. *Nature Physics* **1**, 39-41 (2005).

[6] Emery, N. *et al*. Superconductivity of Bulk CaC$_6$. *Phys. Rev. Lett*. **95**, 087003 (2005).

[7] DiVincenzo, D. P. and Mele, E. J. Self-consistent effective-mass theory for intralayer screening in graphite intercalation compounds. *Phys. Rev. B* **29**, 1685–1694 (1984).

[8] Novoselov, K. S. *et al.* Two-dimensional gas of massless Dirac fermions in graphene. *Nature* **438**, 197-200 (2005).

[9] Zhang, Y., Tan, Y., Stormer, H. L. & Kim, P. Experimental observation of the quantum Hall effect and Berry's

phase in graphene. *Nature* **438**, 201-204 (2005).

[10] Xu, S. *et al.* Energy Dependence of Electron Lifetime in Graphite Observed with Femtosecond Photoemission Spectroscopy. *Phys. Rev. Lett.* **76**, 483-486 (1996).

[11] Moos, G., Gahl, C., Fasel, R., Wolf, M. & Hertel, T. Anisotropy of quasiparticle lifetimes and the role of disorder in graphite from ultrafast time-resolved photoemission spectroscopy. *Phys. Rev. Lett.* **87**, 267402 (2001).

[12] Wallace, P. R. The Band Theory of Graphite. *Phys. Rev.* 71, 622–634 (1947).

[13] Forbeaux, I. Themlin, J. M. & Debever, J. M. Heteroepitaxial graphite on 6H-SiC(0001): Interface formation through conduction-band electronic structure. *Phys. Rev. B* **58**, 16396–16406 (1998).

[14] Shirley, E. L., Terminello, L. J., Santoni, A. & Himpsel, F. J. Brillouin Zone Selection Effects in Graphite Photoelectron Angular Distribution. *Phys. Rev. B* **51,** 13614–13622 (1995).

[15] Bennich, P. *et al.* Photoemission study of K on graphite. *Phys. Rev. B* **59**, 8292–8304 (1999).

[16] Kaminski, A. & Fretwell, H. M. On the extraction of the self-energy from angle-resolved photoemission spectroscopy. *New J. Phys.* **7**, 98 (2005).

[17] Kordyuk, A. A. *et al.* Bare electron dispersion from experiment: Self-consistent self-energy analysis of photoemission data. *Phys. Rev. B* **71**, 214513 (2005)

[18] Valla, T., Fedorov, A. V., Johnson, P. D. & Hulbert, S. L. Many-body effects in angle-resolved photoemission: quasiparticle energy and lifetime of the Mo(110) surface state. *Phys. Rev. Lett.* **83**, 2085–2088 (1999).

[19] Hengsberger, M., Purdie, D., Segovia, P., Garnier, M. & Baer, Y. Photoemission study of a strongly coupled electron phonon system. *Phys. Rev. Lett.* **83**, 592-595 (1999).

[20] Rotenberg, E., Schaefer J., & Kevan, S. D. Coupling between adsorbate vibrations and an electronic surface state. *Phys. Rev. Lett.* **84**, 2925–2928 (2000).

[21] Valla, T. *et al*. Evidence for Quantum Critical Behavior in the Optimally Doped Cuprate Bi2Sr2CaCu2O8. *Science* **285**, 2110-2113 (1999).

[22] Lanzara, A. *et al*. Evidence for ubiquitous strong electron☐phonon coupling in high-temperature superconductors. *Nature* **412**, 510 (2001).

[23] Al-Jishi, R. and Dresselhaus, G. Lattice-dynamical model for graphite. *Phys. Rev. B* 26(8), 4514 (1982).

[24] Vitali, L., Schneider, M. A., Kern, K., Wirtz, L., and Rubio, A. Phonon and plasmon excitation in inelastic electron tunneling spectroscopy of graphite. Phys. Rev. B **69**,121414 (2004).

[25] Grimvall, G. *The Electron Phonon Interaction in Metals*. North Holland Publishing Company, Amsterdam, 1981.

[26] Calandra, M. and Mauri, F. Theoretical Explanation of Superconductivity in $C_6Ca$. *Phys. Rev. Lett.* **95**,237002 (2005).

[27] Kliewer, K. L. and Raether, H. Plasmon Observation Using X Rays, *Phys. Rev. Lett*. **30**(20), 971 (1973).

[28] Jensen, E., Bartynski, R. A., Gustafsson, T., & Plummer, E. W. Distortion of an Unoccupied Band in Be by the Electron-Plasmon Interaction. *Phys. Rev. Lett.* **52**, 2172–2175 (1984).

[29] Li, Z. Y., Hock, K. M., Palmer, R. E. & Annett, J. F. Potassium-adsorption-induced plasmon frequency shift in graphite. *J. Phys.: Condens. Matter* **3,** S103-S106 (1991).